\documentclass[11pt,a4paper]{article}

\pdfoutput=1

\usepackage{jcappub}
\usepackage{mathrsfs}
\usepackage{bbm}
\usepackage{bm}
\usepackage{dsfont}
\usepackage{tensor}
\usepackage{xspace}
\usepackage{aeguill}
\usepackage{wasysym}
\usepackage{microtype}
\usepackage{empheq}
\usepackage{ifthen}
\usepackage{amsmath}
\usepackage{subcaption}
\usepackage{wrapfig}

\renewcommand\lim[2]{\underset{ #1 \rightarrow #2 }{ \mathrm{lim} } \,}

\newcommand{\delimiters}[4][]{
\ifthenelse{ \equal{#1}{1} }{  #2 #3 #4  }
					{ \ifthenelse{\equal{#1}{2}}{ \big#2 #3 \big#4 }
						{ \ifthenelse{\equal{#1}{3}}{ \Big#2 #3 \Big#4 }
							{ \ifthenelse{\equal{#1}{4}}{ \bigg#2 #3 \bigg#4 }
								{ \ifthenelse{\equal{#1}{5}}{ \Bigg#2 #3 \Bigg#4 }
									{ \left#2 #3 \right#4 }
								}
							}
						}
					}
													}

\newcommand{\abs}[2][]{\delimiters[#1]{|}{#2}{|}}

\newcommand{\mean}[2][]{\delimiters[#1]{\langle}{#2}{\rangle}}

\newlength{\boxtitlelength}
\newlength{\halfrulelength}
\newcommand{\boxtitle}[1]{\footnotesize\bf{\:#1\:}}

\definecolor{blue4}{RGB}{0,0,143}
\definecolor{red4}{RGB}{143,0,0}
\definecolor{orange}{RGB}{255,128,0}
\definecolor{darkcyan}{RGB}{0,128,128}
\definecolor{olive}{RGB}{0,128,0}
\definecolor{purple}{RGB}{128,0,128}
\definecolor{cyan2}{RGB}{0,255,255}
\definecolor{fushia}{RGB}{255,0,255}
\definecolor{mygray}{gray}{0.5}
\definecolor{lightgray}{gray}{0.85}

\newcommand{\apj}{ApJ}

\newcommand{\mnras}{MNRAS}

\newcommand{\jcap}{JCAP}

\newcommand{\prd}{Phys. Rev. D}

\usepackage{amssymb}
\usepackage{booktabs}  % Enhances the quality of tables in LaTeX
\usepackage{dcolumn}  % Align table columns on decimal point
\usepackage{graphicx}

\newcommand{\be}{\begin{equation}}
\newcommand{\ee}{\end{equation}}
\newcommand{\kg}{{k_\mathrm{g}}}
\newcommand{\kd}{{k_\mathrm{d}}}
\newcommand{\post}[3]{P(#1 \mid #2, #3)}
\newcommand{\like}[3]{\mathcal{L}(#1 \mid #2, #3)}
\newcommand{\prior}[2]{\mathcal{P}(#1 \mid #2)}
\newcommand{\evid}[2]{\mathcal{E}(#1 \mid #2)}
\newcommand{\dif}[1]{\mathrm{d}#1}
\newcommand{\N}[2]{\mathcal{N} \left( #1, #2 \right)}
\newcommand{\U}[2]{\mathcal{U} \left( #1, #2 \right)}

\newcommand{\aanda}{Astron. Astrophys.}
\newcommand{\aipcp}{AIP Conf. Proc.}
\newcommand{\cqg}{Class. Quant. Grav.}
\newcommand{\ijmpd}{Int. J. Mod. Phys. D}
\newcommand{\prl}{Phys. Rev. Lett.}

\newcolumntype{d}[1]{D{.}{.}{#1}}  % To align to the dot
\newcolumntype{v}[1]{D{,}{,\ }{#1}}  % To align to the comma
\newcolumntype{p}[1]{D{,}{\ \pm\ }{#1}}  % To align to the plus-minus signal

\def\mean#1{{\vphantom{\tilde#1}\bar#1}}
\def\bH{\mean H}\def\OM{\mean\Omega}

\def\ns#1{_{\hbox{\sevenrm #1}}} 
\def\goesas{\mathop{\sim}\limits} 
 
\def\Y#1{^{\raise2pt\hbox{$\scriptstyle#1$}}}
\def\Z#1{_{\lower2pt\hbox{$\scriptstyle#1$}}}
\def\X#1{_{\lower2pt\hbox{$\scriptscriptstyle#1$}}}
 
\def\QQ{{\cal Q}} 
 \def\kv{k\ns v} 
 
 \def\fvi{{f\ns{vi}}}

 \def\fv{{f\ns v}}  
  
\def\OMk{\OM\Z k}\def\OMQ{\OM\Z{\QQ}}

 \def\Hb{\bH\Z{\!0}} \def\Hh{H} \def\Hm{H\Z0}
\font\sevenrm=cmr7

\def\lsim{\mathop{\hbox{${\lower3.8pt\hbox{$<$}}\atop{\raise0.2pt\hbox{$\sim$}}
$}}}

\def\fvn{f\ns{v0}}  

\newcommand{\lcdm}{$\Lambda$CDM\ }
\newcommand{\OmegaMo}{\Omega_{\mathrm{m},0}}
\newcommand{\OmegaBo}{\Omega_{\mathrm{b},0}}
\newcommand{\OmegaRo}{\Omega_{\mathrm{r},0}}
\newcommand{\OmegaKo}{\Omega_{k,0}}
\newcommand{\OmegaKdo}{\Omega_{k_\mathrm{d},0}}
\newcommand{\OmegaKgo}{\Omega_{k_\mathrm{g},0}}

\newcommand{\zd}{{z_\mathrm{d}}}
\newcommand{\Neff}{{N_\mathrm{eff}}}

\makeatletter
\def\@fpheader{\relax}
\makeatother

\title{Data Analysis and Phenomenological Cosmology}

\author[a]{Alan A. Coley,}
\author[b]{Beethoven Santos,}
\author[a]{Viraj A. A. Sanghai}

\affiliation[a]{School of Mathematics \& Statistics, Dalhousie University,\,\\
6316 Coburg Road, Halifax, B3H 4R2, Canada.}
\affiliation[b]{Coordenação de Astronomia e Astrofísica, Observatório Nacional, 20921-400,\\Rio de Janeiro -- RJ, Brasil.}

\emailAdd{aac@mathstat.dal.ca}
\emailAdd{thoven@on.br}
\emailAdd{viraj.a.a.sanghai@dal.ca}

\abstract{
    In the era of precision cosmology, even percentage level effects are significant on cosmological observables. The recent tension between the local and global values of $H_0$ is much more significant than this, and any possible solution might rely on us going beyond the standard \lcdm cosmological model. For much smaller, yet potentially significant effects, spatial curvature from averaging and cosmological backreaction on observational predictions could play a role. This is especially true with the higher precision of new observational data and improved statistical techniques. In this paper, we discuss the observational viability of a class of physically motivated cosmologies which can be parametrized by a phenomenological two-scale backreaction model with decoupled spatial curvature parameters and two Hubble scales. Using the latest JLA Supernovae data together with some of the latest BAO data, we perform a Bayesian model selection analysis and find that the phenomenological models are not favoured over the standard \lcdm cosmological model. Although there is still a preference for non-zero and unequal dynamic and geometric spatial curvatures, there is little evidence for differing Hubble scales within these phenomenological template models.
}


\date{\today}

\arxivnumber{1808.07145}
\begin{document}

\maketitle
\flushbottom

\section{Introduction}
\label{sec:introduction}

In the standard \lcdm cosmological model, the Universe is described by a Friedmann--Lemaître--Robertson--Walker (FLRW) geometry satisfying Einstein's field equations (EFE) of general relativity (GR), with a cosmological constant, $\Lambda$. The FLRW model can perhaps be interpreted as a phenomenological template to compare cosmology with observations \cite{review}, since there are issues regarding its physical motivation and underpinning \cite{NotGW}. It is possible that it would be useful to replace the
phenomenological FLRW cosmological equations with another extended set of phenomenological equations. For example, in an inhomogeneous universe, while gravity is governed by EFE on small scales, this is not necessarily the case for large-scale averages. Consequently, an important problem in cosmology is to determine the precise size and form of deviations from EFE when considering averaged geometric effects on large scales, and what effects such ``backreactions'' will have on observations \cite{review}. Most authors accept that the effects of backreactions will be important for precision cosmology \cite{NotGW, Sanghai:2017yyn}.

In addition, due to the abundance of new observational data \cite{Betoule2014, Wang2017, Carvalho2016, Alcaniz2017} and improved statistical techniques, it is perhaps timely to further study the constraints on cosmological models. In the analysis of standard cosmological models some parameters are purely phenomenological. Therefore, we study a parametrized phenomenological model (which has phenomenological constants), within which we can analyze data, in order to study the potential observational effects of more general cosmological models. If there is any observational evidence that the phenomenological parameters do not to take on their standard conventional values in these more general models, then it will motivate further study of these models.

Cosmological observations probe the Universe on different scales; high redshift type Ia supernova (SNe Ia) cover a range of scales out to several Gpc, while the Cosmic Microwave Background (CMB) involves making observations on the scale of the cosmological horizon ($\sim14$\,Gpc). The motivation to study an phenomenological extension to the standard model is two-fold: i) There is a $3.3\sigma$ tension between the local Hubble constant from type 1a supernova and the large-scale Hubble constant obtained from the CMB \cite{R16, planck2015, R18, planck2018}. ii) In previous works \cite{SantosColey, CCCS}, using older Baryonic Acoustic Oscillation (BAO) data \cite{Beutler2011, Ross2015, Anderson2014, Padmanabhan2012, Blake2012}, it has been shown that a simple two-curvature extension to the standard model might be slightly preferred over the \lcdm model.

The most recent determination of the local value of the Hubble constant based on direct measurements made with the Hubble Space Telescope \cite{R18}, of $H_0=73.52\pm1.62$ km/s/Mpc, is now more than $3$ standard deviations higher than the value derived from the most recent CMB anisotropy data provided by the Planck satellite in a \lcdm model of $H_0=67.27\pm0.60$ km/s/Mpc \cite{planck2018}. This motivates the simple phenomenological model with different Hubble parameters at different scales which we study here.

In addition, \cite{SantosColey} studied the observational viability of a simple phenomenological two-scale model with a simple parametrized backreaction contribution to the EFE with non-equal spatial curvature parameters $\kg$ and $\kd$ \cite{CCCS}, motivated by an exact and fully covariant macroscopic averaging procedure~\cite{Coley:2005ei}. Current constraints on spatial curvature within the standard \lcdm model show that it is dynamically negligible: $\Omega_k \sim 5\times 10^{-3} $ (95\%CL) \cite{planck2015}. However, using recent SNe Ia data and measurements of the BAO scale, it was found that the combination of these data sets suggest unequal values of the two spatial curvature parameters, with the constraints on the spatial curvature parameter being significantly weaker than those in the standard model. Such a relatively large positive detection of spatial curvature, together with evidence of scale dependence of the spatial curvature, is a sure sign of non-trivial averaging effects \cite{Coley:2005ei}. In addition, such a value cannot be naturally explained by inflationary models that allow a large number of e-foldings which predict that the effective curvature within our Hubble radius should be of the order of the amplitude of the curvature fluctuations generated during inflation, i.e. $\Omega_k \sim 10^{-5}$ \cite{DiDio}. Note, however, that an open Universe is strongly suggested by the string multiverse \cite{Leonard}.

In section \ref{back}, we begin by briefly recapping the tension in the Hubble constant and the proposals that have been put forward to alleviate this tension. We then briefly discuss the work done to motivate the study of spatial curvature in cosmology. Also, we recap the standard Bayesian statistical methods used to estimate the standard cosmological parameters. In section \ref{exten}, we introduce a parametrized phenomenological model that is similar to the \lcdm model but, in general, contains two-Hubble constants and two-spatial curvatures, both of which depend on scale. We give examples of three physically motivated models that fit within this framework -- the two curvature model, the two-scale fractal bubble model and the the bi-domain scalar averaging model. In section \ref{method}, we elaborate on the method that we use to perform our statistical analysis, including our choice of priors. Finally, in section \ref{results}, we present the results we've obtained, before we discuss the implications of these results.

\section{Background} \label{back}

\subsection{Tension in Hubble constant}

Since the 2013 data release of the Planck data, the constraints on the Hubble constant coming from the Planck satellite have been in significant tension with the 2011 supernova data, based on direct measurements made with the Hubble Space Telescope. This tension was further confirmed in the 2015 Planck data release; assuming standard \lcdm the Planck data gives $H_0=67.51\pm0.64$ km/s/Mpc \cite{planck2015}. Although a large number of authors have proposed several different mechanisms to explain this tension, after three years of improved analyses and data sets, the tension in the Hubble constant between the various cosmological datasets not only persists but is even more statistically significant. The recent analysis of \cite{R16} found no compelling arguments to question the validity of the dataset used. %(which had been questioned by some authors).
%At the same time, the new constraints on the reionization optical depth, obtained with Planck High Frequency Instrument (HFI) data, bring the Planck constraint on $H_0$ to an even lower value, with $H_0=66.93\pm0.62$ km/s/Mpc at $68 \%$ CL \cite{planckHFI} (although the presence of systematics is not yet excluded).
Indeed, the recent determination of the local value of the Hubble constant by Riess et al, 2016 \cite{R16} (hereafter R16) $H_0=73.24\pm1.74$ km/s/Mpc at $68 \%$ confidence level (CL) is now about $3$ standard deviations higher than the (global) value derived from the 2015 CMB anisotropy data provided by the Planck satellite assuming a \lcdm model. This tension only gets worse when we compare the Riess et al, 2018 $H_{0}$ value of $H_0=73.52\pm1.62$ km/s/Mpc \cite{R18} to the Planck 2018 $H_{0}$ value of $H_0=67.27\pm0.60$ km/s/Mpc \cite{planck2018}. The Riess et al, 2018 results were obtained by using additional Galactic Cepheids data and Gaia parallaxes, whereas the Planck 2018 analysis includes the CMB temperature, polarization and lensing data. These observations allowed us to obtain the latest local and global constraints on $H_{0}$.

In order to investigate possible solutions to the Hubble constant tension a number of proposals have been made \cite{VMS1, VMS3, VMS4, Sola1}. A combined analysis of the Planck and R16 results, in an extended parameter space, was performed in \cite{VMS1}. They simultaneously varied $12$ cosmological parameters instead of the usual $6$ in \lcdm \cite{VMS2}. The additional parameters that were considered were the dark energy equation of state, the neutrino effective number, the running of the spectral index, the tensor to scalar ratio, the neutrino mass and, finally, the purely phenomenological amplitude of the gravitational lensing on the CMB angular spectra (which comes from the Planck data itself). It was found that a phantom-like dark energy component, with effective equation of state $w=-1.29_{-0.12}^{+0.15}$ at $68 \%$ CL \cite{VMS1} can solve the current tension between the Planck dataset and the R16 results in an extended \lcdm scenario. This result was also confirmed when including cosmic shear data from the CFHTLenS survey and CMB lensing constraints from Planck. However, when BAO measurements are included it was found that some of the tension with R16 remains, as is also the case when we include the supernova type Ia luminosity distances from the Joint Light-curve Analysis (JLA) catalog.

More recently, an extended parameter space was used to show how interacting dark-energy models could also alleviate the $H_{0}$ tension \cite{VMS3}. Along with the interacting dark energy, a phantom-like dark energy with effective equation of state $w=-1.184\pm 0.064$ at $68 \%$ c.l \cite{VMS3} instead of a cosmological constant with $w=-1$, provided a better fit to the data, and also alleviated the $H_{0}$ tension. However, when the Planck data was combined with external data sets such as BAO, JLA type Ia supernova, cosmic shear or lensing data, no compelling evidence for interacting dark energy was found \cite{VMS3}. A vacuum phase transition model, physically motivated by quantum gravity effects, and with the same number of parameters as $\Lambda$CDM, was also shown to alleviate the $H_{0}$ tension \cite{VMS4}. However, external datasets, such as BAO data, were not applied to this analysis. Other dynamical dark energy models, such running vacuum models \cite{Sola1}, have also been shown to alleviate the $H_{0}$ tension and favour lower $H_{0}$ values of Planck data rather than the local $H_{0}$ values of R16 data.

All of the models provided here have used a phenomenological template to alleviate the $H_{0}$ tension, motivated by physical models that fit within this template. We will do something similar here, albeit with a slightly different phenomenological template and physical motivation.

Different methods to improve upon the uncertainties and systematics for R16 have also been employed to help alleviate the $H_{0}$ tension. In \cite{Cardona1}, Cepheids data was re-analysed using Bayesian hyper-parameters to take into account the uncertainties in a more conservative way. However, this did not get rid of the $H_{0}$ tension with the Planck data and if anything, further increased it.

%{\bf{EDIT MORE AND UPDATE: VIRAJ}}

\subsection{Spatial curvature in cosmology}

Recent work, and especially that of \cite{SantosColey}, are timely and motivate models with curvature at the level of a few percent. The effect of non-linear structure on the background cosmological expansion together with the averaging process used to understanding this effect is referred to as ``backreaction'' in cosmology. Backreaction estimates tend to give {a strong} negative mean curvature; indeed, the volume averaging of the 3-spatial curvature over a volume occupied by the remaining non-virialized matter is expected to be negative \cite{Larena09template}. Also, contrary to the popular belief that spatial curvature is severely observationly constrained, observations on recently emerged, {\em present-day} {(large-scale mean)} average negative curvature are weak and not easy to measure \cite{Larena09template}, but will become measurable in near-future surveys such as Euclid. In addition, the current curvature parameter estimations are not yet at the cosmic variance limit (beyond which constraints cannot be meaningfully improved due to the cosmic variance of horizon scale perturbations); indeed, the current measurements are more than one order of magnitude away from the limiting threshold \cite{DiDio, Leonard}.

We note that an observation of non-zero spatial curvature could be the result of large-scale structure effects. Local inhomogeneities and perturbations to the distance-redshift relation at second-order contribute a monopole at the sub-percent level, leading to a shift in the apparent value of $\Omega_k$. Other GR effects related to the behaviour of curvature in inhomogeneous spacetimes also contribute an apparent shift in the background cosmology. Given that current curvature upper limits are 1 or 2 orders of magnitude away from the level required to probe most of these effects, there is an imperative to continue pushing $\Omega_k$ constraints to greater precision (i.e., to about 0.01\% level) \cite{Leonard}. In addition, current standard measurements of the curvature parameter come from the combination of CMB and LSS, which does not take into account certain GR effects. In an investigation of how future measurements of the spatial curvature parameter $\Omega_k$ are affected by GR effects, it was shown that constraints on the curvature parameter may be strongly biased if
cosmic magnification is not included in the analysis \cite{DiDio}. Indeed, assuming flatness exclusively would preclude a number of potentially powerful tests of early Universe physics \cite{Leonard}, and of GR effects in large-scale structure \cite{DiDio}. In particular, to test models of inflation such as eternal inflation, we need to reach the cosmic variance limit for spatial curvature \cite{Leonard}.

Current constraints on curvature from Planck 2015 are $|\Omega_k|< 5 \times 10^{-3}$ at 95\% CL \cite{planck2015}. Note that the Planck 2018 results do not improve on this with constraints of $|\Omega_k|< 10^{-2}$ at 95\% CL \cite{planck2018}. However, another factor that we must consider while quoting constraints on spatial curvature in cosmology is this usually relies upon strong assumptions on the form of dark energy in our model \cite{Witzemann}. More recently there have been works that have tried to constrain spatial curvature in a model independent manner (without a dependence on the form of dark energy). With the current experimental setups as they are, it's not quite clear how we would reach the cosmic variance limit in such a situation \cite{Witzemann}.

Another interesting idea was how emergent spatial curvature might be able to account for the discrepancy between the local and global values of $H_{0}$ \cite{Bolejko1}. The idea was based on a model in which you start off with a flat \lcdm model + perturbations at early times and then evolve it forward, where non-linear structures result in a slightly curved universe \cite{Bolejko2}. Performing a ray-tracing simulation in such a universe results in a slightly lower value of $H_{0}$ at higher redshifts and a higher value of $H_{0}$ at lower redshifts. The values of $H_{0}$ obtained are in line with those expected from observations.

%{\bf{UPDATE: VIRAJ}}

\subsection{Data analysis in standard model}

There are alternatives to the \lcdm model that have not yet been ruled out by observations. Constructing a viable physical model is not just a question of fitting the model to the data, it is also a question of model selection. This necessitates robust statistical techniques that allow for reasonable decisions using incomplete information. Bayesian inference provides a quantitative framework for plausible conclusions \cite{DebonoSmoot}.

At the initial level of Bayesian inference, we can estimate the allowed parameter values of the theory. If~we assume a \lcdm universe within GR, we still need to fix the values of the various~constants (e.g., a cosmological constant or a dynamical dark energy~parameter such as a dark energy equation of state parameter $w$). Next, we can ask which parameters we should include in the theory. Although the current data is effectively consistent with a six-parameter \lcdm model based on GR, there are many more possible parameters which could be included in future data analysis. It is not viable to simply include all possible parameters to fit the data, since each may give rise to degeneracies that weaken constraints on other parameters (including the \lcdm parameter set). If we relax the assumption that the theory of gravity is GR, then many more parameters are possible \cite{Trotta2008}.

If the goal of the data analysis is to decide which parameters need to be included in order to explain the data, then those extra parameters need to be physically motivated and the physical effects to which the data are sensitive determined. At the current time, alternative models are viable, and we therefore require a consistent method to discard or include parameters. This is the second level of Bayesian inference---model selection. Bayesian model selection penalizes models which introduce unnecessary parameters. A theory can always be constructed that fits the data perfectly, even better than GR, but we might need to introduce extra free parameters. Consequently a balance between goodness of fit (the degree of complexity) and predictive power (consistency with prior knowledge) is sought. Possibly GR and the \lcdm model fits all the data with the minimum number of parameters. Recently cosmologists have started to use Bayesian methods for cosmological model selection in relation to GR, since the observational data has begun to have the necessary statistical power to facilitate model testing \cite{Trotta2008}.

We should also note that the data has often been analyzed using statistics specifically developed for FLRW models, which are not necessarily reliable (or optimal) for models not close to FLRW. Better statistical fits are likely probable utilizing an appropriately adapted framework.

%{\bf{OKAY? MORE?}}

\section{Extended phenomenological model} \label{exten}

We present a parametrized phenomenological model, similar in spirit to other phenomenological extensions of the standard model \cite{VMS1}, within which we can statistically analyze data in order to study the potential observational consequences of non-trivial phenomenological parameters.
There are a number of existing implementations of emerging average negative curvature models, in addition to the phenomenological two-curvature model which parametrizes possible backreaction contribution, including the more physical bi-scale fractal bubble model \cite{opus}, based on the timescape model \cite{Wiltshire09timescape} and the bi-domain scalar averaging approach based on the template model \cite{Larena09template} and the virialisation approximation \cite{ROB13}, which can be utilized to study the potential observational effects of averaging.

The metric of the local spacetime geometry in the phenomenological models is assumed to be given by:
\be
    \label{Pmetric}
    ds^2 = -dt^2 + a^2(t) \left[ \frac{dr^2}{1-\kg r^2} + r^2 d\Omega \right],
\ee
with local geometrical curvature, $\kg$, with ${\Omega}_{\kg} \equiv - {\kg}/{a_0^2 H_0^2}$, and scale factor $a(t)$ with corresponding Hubble parameter $H$.

On large scales the macroscopic FE are:
\be
    \label{PFE}
    \bar{H}^2= \alpha^2 H^2= \frac{8 \pi G}{3} \bar{\rho} -\frac{\kd}{\bar{a}^2} + \frac{\Lambda}{3},
\ee
where the `dynamical curvature', $\kd$, includes contributions from both the spatial curvature and correlations (backreaction) and, in general, in spacetimes that are inhomogeneous on small scales, we do not expect the two `spatial curvature' terms $\kg,\kd$ to be equal when the backreaction cannot be neglected.
In addition, the Hubble parameter $\bar{H}= H_d$ derived from the effective scale factor $\bar{a}$ is not equal to the Hubble parameter ${H}= H_g$, and $\alpha$ represents a new phenomenological parameter, $ \bar{H}\equiv \alpha H$, characterizing the two Hubble scales.
Note that ${\bar{a}} = a^{\alpha}$, where $\bar{a}(t_0) = a(t_0) =1.$

We define the normalized constants:
\be
    \label{PFE3}
    \bar{\Omega}_{m} = \frac{8 \pi G \bar{\rho}_0}{3 \alpha^2 a_0^2 H_0^2}, ~~
    \bar{\Omega}_{\kd} = - \frac{{\kd}}{\alpha^2 a_0^2 H_0^2}, ~~
    \bar{\Omega}_{\Lambda} = \frac{{\Lambda}}{\alpha^2 a_0^2 H_0^2}
\ee
(we could absorb $\alpha$ by renormalizations and redefining the curvature constants) the macroscopic Friedmann equation again becomes at $t = t_0$
\be
    \label{PFE2}
    \bar{\Omega}_m + \bar{\Omega}_{\kd} + \bar{\Omega}_{\Lambda}=1.
\ee
In this model the spatial curvature of the spacetime metric is decoupled from the spatial curvature that appears in the macroscopic Friedmann equation \cite{CCCS}, as are the Hubble constants $\bar{H}$ and $H$ when $\alpha \neq 1$.
We refer to this as the ``2CC-Ext'' model.

The trajectories of (the average of a large number of) photons are null trajectories with respect to the spacetime metric.
Integrating then gives rise to the luminosity distance-redshift relation:
\be
    \label{PdA}
    d_L (z) = \frac{(1+z)}{H_0 \sqrt{\vert \Omega_{\kg} \vert}}
              f_\kg \left(
                  \int_{\frac{1}{1+z}}^{1}
                      \frac{\sqrt{\vert {\Omega}_{\kg} \vert} d\tilde{a}}{\sqrt{\bar{\Omega}_{\kd} \tilde{a}^{4-2\alpha} + \bar{\Omega}_{\Lambda} \tilde{a}^4 + \bar{\Omega}_m \tilde{a}^{4-3\alpha} }}
                \right),
\ee
where $f_\kg(x)=\sinh (x)$, $x$ or $\sin (x)$ when $\kg <0$, $\kg=0$ or $\kg>0$, respectively.
Note that this model reduces to the standard model when ${\kd}=\kg$ and $\alpha =1$.

\subsection{Two curvature model}

The phenomenological two-scale model with a simple parametrized backreaction contribution to the FE, motivated by an exact and fully covariant macroscopic averaging procedure \cite{Coley:2005ei}, and which has decoupled spatial curvature parameters in the metric and the Friedmann equation and reduce to the FLRW solutions of EFE when these parameters are equal, was studied in \cite{CCCS}.

The metric of the spacetime geometry in the phenomenological two curvature cosmological model is given by {equation} (\ref{Pmetric}) with geometrical curvature, $\kg$, and scale factor $a(t)$.
On large scales the macroscopic FE are given by {equation} (\ref{PFE}), which leads to eqn. (\ref{PFE2}), where the `dynamical curvature', $\kd$, includes contributions from both the spatial curvature and correlations (backreaction).
In general, in spacetimes that are inhomogeneous on small scales, we do not expect these two `spatial curvature' terms to be equal.
However, in this model we have that $\alpha=1$, so that $\bar{H}=H$.
We refer to this as the ``2CC'' model.
The luminosity distance-redshift relation is given by {equation} (\ref{PdA}).

Constraints on the phenomenological two curvature models were first investigated in \cite{CCCS}, and improved upon by \cite{SantosColey}.
It was found that the additional freedom gained by allowing $\Omega_{\kg} \neq \Omega_{\kd}$ is considerable, with constraints on $\Omega_{\Lambda}$ and the two $\Omega_{k}$'s being significantly weaker than in the standard approach.
The combination of all of these observables still appears to provide strong evidence for the existence of dark energy.
However, it is striking that constraints on $\Omega_{\kg}$ are an order of magnitude tighter than those on $\Omega_{\kd}$.
There are even tantalizing hints that the data may favor $\Omega_{\kg} \neq \Omega_{\kd}$ (at a high level of confidence; see, for example, \cite{CCCS}).
In addition, in \cite{SantosColey} a Bayesian model selection statistics to compare the predictivity power of the averaged model with respect to the standard \lcdm cosmology was also performed, and it was found that although the current SNe Ia data alone cannot distinguish between these two models, the combination of SNe Ia and BAO data sets significantly favors significantly the two curvature model with respect to the standard \lcdm cosmology.

\subsection{Two--scale, fractal bubble model}

The two--scale fractal bubble model or timescape model was presented in \cite{opus,Wiltshire09timescape} and analysed in \cite{paper2}.
Two sets of cosmological parameters are relevant: those relative to an ideal observer at the volume--average position in freely expanding spacetime and local conventional dressed parameters.
The conventional metric arises in our attempt to fit a single global metric to the universe with the assumption that average spatial curvature and local clock rates everywhere are identical to our own, which is no longer true.

The Buchert {\bf{volume averaged}} quantities (as measured with canonical clocks) with statistically averaged (or bare) Hubble parameter $\bar{H}$ is applicable on large scales (scales larger than the course-graining scale $100/L$; valid for large redshifts of $z\goesas37$) and is appropriate for analysis of CMB observations at the the surface of last scattering \cite{opus}.
The averaged spatial curvature is assumed to be negative, due to the predominance of voids at larger scales.
The volume--averaged matter, curvature and kinematic back-reaction variables are denoted by $\bar{\Omega}_m, \OMk, \OMQ$, respectively, and given in terms of the parameters $\kv, \fvi, \fv$, where $\bar{\Omega}_m + \bar{\Omega}_{{\cal X}}+\bar{\Omega}_\Lambda=1$ and $\bar{\Omega}_{{\cal X}} \equiv \OMk+\OMQ$.
At late times and at large scales the quantities $\bar{H}_0$ and $k_d$ can be approximated as constants, and the metric can be approximated by an FLRW model with negative constant spatial curvature: $\bar{\Omega}_{{\cal X}}=\bar{\Omega}_{k_d}$.

The {\bf{dressed geometry}} is assumed to be approximately spherically symmetric and does not necessarily have constant spatial curvature everywhere.
For example, the `locally' measured dressed Hubble parameter $H$ is that inferred by a local wall observer (such as ourselves) trying to fit an ideal FLRW geometry, and is appropriate for smaller scales ($z>0.03$) and suitable for analysing Supernova data.
The local effective average geometry at the boundary of a finite infinity region is assumed to be essentially spatially flat with the metric in terms of clocks and rulers as measured by a local wall observer.
The conventional dressed matter density parameter, $\Omega_m$, is expected to take numerical values similar to those we infer in FLRW models, but differs from the bare volume--average density parameter, $\Omega_m$.
The dressed Hubble parameter that we measure as wall observers, the global average over both walls and voids, is not $\bH$, but $\Hh$.
The dressed Hubble constant, $\Hm$, is then related to the bare Hubble constant, $\Hb$, by $\Hb = \alpha\Hm$, where ${1/\alpha} \equiv(4\fvn^2+\fvn+4)/2(2+\fvn)$ in terms of the parameters of the model, and we expect that $\Hm>\Hb$ so that the ratio of the Hubble scales $\alpha<1$.

At late times, and utilizing the exact special two--scale tracker solution of the Buchert equations \cite{buch1}, it follows that $\alpha$ (and $k_g$), is approximately constant when evaluated at the present epoch and appropriate for comparing the model with observations.
We stress that by taking the variables as constants in a phenomenological model we hope to capture some of the essential features of the physically motivated models that can be tested, and if they are found to take on non-standard values this would then motivate further study of the more physical models.

\subsection{The bi-domain scalar averaging approach}

In inhomogeneous GR cosmology, structure formation couples to average cosmological expansion.
In a conservative bi-domain scalar averaging approach, an Einstein--de~Sitter model (EdS) is assumed at early times and extrapolated forward in cosmological time as a ``background model'' against which the average properties of today's Universe are measured \cite{RTBO}.
Space--time is foliated using a 3+1 ADM procedure with metric in local synchronous coordinates.
The Friedmann and Raychaudhuri equations are extended from the spatially homogeneous case in GR to take into account inhomogeneous curvature and inhomogeneous expansion of the Universe.
Spatial slices are divided into into overdense (virialised collapsed) regions and underdense (void) regions \cite{ROB13}.
An effective averaged model is then built by the scalar averaging procedure of \cite{buch1}.

In order to test whether the averaged inhomogeneous cosmologies can correctly describe observations of the large scale properties of the current Universe, a smoothed {\em template metric} is introduced which corresponds to a constant spatial curvature FLRW model at any time, but with an evolving scalar curvature parameter related to the true averaged scalar curvature.
Photons move on null-geodesics of this template metric, which is used to compute quantities along an approximate effective lightcone of the averaged model of the Universe (which cannot be simply approximated by a FLRW lightcone).
The standard \lcdm case can be recovered by formally taking $\Omega_{{\cal X}} = \Omega_k$ on a chosen large domain.

At early epochs, prior to the main virialisation epoch, the expansion is dominated by the EdS background model with present Hubble parameter ${H_0}$,
while at the present, the effective local expansion as measured by local estimates of the (effective low-redshift) Hubble constant ${\bar{H}_0}$, is the sum of the background expansion rate and the peculiar expansion rate of voids, $H_{0}^{{void}}$; i.e., $\bar{H}_0 \approx {H_0} + H_{0}^{{void}}$.
For a fixed large scale of statistical homogeneity the dynamics is dominated by the underdense regions, especially at late times.
Physically we have positive ${H_0}$.
We have a void-dominated model, so we also have positive ${H_{0}^{{void}}}$.
Thus, the solution of physical interest is of the form ${H_0} \propto {\bar{H}_0}$ at high-redshift ($z > 3$).
The ratio ${H_0}/{\bar{H}_0}$ can be obtained in a particular cosmological model, such as a template metric parameterized by exact scaling properties of an averaged inhomogeneous cosmology, and might be as small as ${H_0}/{\bar{H}_0} \sim 1/2$ \cite{ROB13}.

In \cite{Larena09template} a likelihood analysis was performed to infer constraints on backreaction models using the SN Ia data from the SNLS collaboration and the position of the CMB peaks and dips from WMAP3--yr data measured.
The effects of backreaction can be estimated in a simple class of scaling solutions.
For example, with the constant (present-day) $\Omega_{{\cal X}} \approx 0.7$, in a constant cosmological time, constant-curvature hypersurface at $z=1$, a subpercent (tangential stretching) effect occurs at 500 $h^{-1}$ Mpc from the observer, while at the BAO scale of about $105$ $h^{-1}$ Mpc, the effect is considerably weaker \cite{Larena09template}.
Current data do not disfavour such a backreaction model.
It is anticipated that the experimental uncertainties on the angular diameter distance and the Hubble parameter from BAO measurements -- forseen in future surveys like the proposed EUCLID satellite project -- are sufficiently small to distinguish between a FLRW template geometry and the template geometry with consistently evolving curvature.

\section{Method and Analysis} \label{method}

We use a similar method to that in \cite{SantosColey}. In this section, we apply Bayesian statistics to perform a parameter estimation for the 2CC-Ext model. Also, we compute the Bayesian Evidence to perform a model comparison between the 2CC-Ext model and \lcdm scenarios.

\subsection{\label{subsec:estimation}Parameter estimation}

Using Bayesian inference, we analyze our dataset $D$ by computing the joint posterior $P$. For a set $\Theta$ of free parameters, we use Bayes' Theorem~\cite{Bayes1764}:
\be
    \label{eq:BayesTheorem}
    \post{\Theta}{D}{M} = \dfrac{\like{D}{\Theta}{M} \, \prior{\Theta}{M}}{\evid{D}{M}} \,,
\ee
where $M$ is the \emph{model}, $P$ is the \emph{posterior}, $\mathcal{L}$ is the \emph{likelihood}, $\mathcal{P}$ is the \emph{prior} and $\mathcal{E}$ is the \emph{Evidence}, where we have suppressed their arguments. For a subset of physically interesting parameters $\theta$ and a subset of nuisance parameters $\phi$, the full set of parameters is $\Theta = (\theta, \phi)$. Hence, we can write the posterior~(\ref{eq:BayesTheorem}) on the parameter of interest marginalized over the nuisance parameters as
\be
    \label{eq:BT_parameter}
    \post{\theta}{D}{M} \propto \int \like{D}{\Theta}{M} \, \prior{\Theta}{M} \, \dif{\phi} \,,
\ee
where the proportionality symbol ``$\propto$'' is due to the fact that the Evidence in~(\ref{eq:BayesTheorem}) is a normalization constant. This makes the Evidence irrelevant in parameter estimation.

Modern methods to compute the posterior in~(\ref{eq:BayesTheorem}) rely on Markov Chain Monte Carlo (MCMC) sampling techniques. Some standard MCMC algorithms can be found in refs.~\cite{Metropolis1953, Mackay2003, Skilling2004, Feroz2013} and its applications to cosmology can be found in refs.~\cite{Lewis2002, Mukherjee2006}.
%\textbf{
Here we use the \textsc{MultiNest}\footnote{\url{https://ccpforge.cse.rl.ac.uk/gf/project/multinest}} algorithm~\cite{Feroz2008, Feroz2009, Feroz2013} through its Python package \texttt{PyMultiNest}\footnote{\url{https://johannesbuchner.github.io/PyMultiNest}}~\cite{Buchner2014}.
%}
Algorithms such as these are used for fast computations of likelihoods and sampling a large number of different kinds of distributions.

\subsection{\label{subsec:lcdm_comp}Comparison to \lcdm}

As stated above, we need to compute the Bayesian Evidence $\mathcal{E}$ in~(\ref{eq:BayesTheorem}) to perform a model comparison. Given a data set $D$ and a given model $M$, we can evaluate the probability of a given model over the full parametric space of the model so that:
% It evaluates the model's performance in the light of the data by integrating the product over the full parametric space of the model:
\be
    \label{eq:evidence}
    \evid{D}{M} = \int_M \like{D}{\Theta}{M} \, \prior{\Theta}{M} \, \dif{\Theta} \,,
\ee
where we integrate over the product of the likelihood $\mathcal{L}$ and prior $\mathcal{P}$. Numerically and computationally, integrating~(\ref{eq:evidence}) can be very difficult.
%\textbf{
To compute the evidence values we use the \texttt{MultiNest} algorithm, requiring a global log-evidence tolerance of 0.01 as a convergence criterion and working with a set of 1000 live points to improve the accuracy in the estimate of the evidence. With this number of live points, the number of samples for all posterior distributions was of order $\mathcal{O}(10^4)$.
%}

% Bayes Factor
We discriminate between the 2CC-Ext model and the standard \lcdm model by using the Bayes factor. The Bayes factor, $B$, is defined as $B \equiv \mathcal{E}_\mathrm{2CC-Ext} / \mathcal{E}_{\Lambda\text{CDM}} $, where $\mathcal{E}_{\Lambda\text{CDM}}$ is the Bayesian Evidence of the \lcdm model and $\mathcal{E}_\mathrm{2CC-Ext}$ is the Bayesian Evidence of the extended phenomenological model. To interpret the values of the Bayes factor, we use a slightly conservative version of the so-called Jeffreys' Scale~\cite{Jeffreys1961}. Values of $\abs{\ln{B}}< 1$ indicate an \emph{inconclusive} evidence whereas values of ${\ln{B}}$ above 1, 2.5 and 5 indicate a \emph{weak}, \emph{moderate} and \emph{strong} evidence in favour of the 2CC-Ext model, respectively. If $\ln{B} < -1$, it means there is support in favour of the standard model.

%To discriminate scenario, we computed the Bayes Factor of the \lcdm model relative to the 2CC model, given by $B \equiv \mathcal{E}_{\Lambda\text{CDM}} / \mathcal{E}_\mathrm{2CC}$, and adopted the following scale to interpret the values of $\ln{B}$: values of $\abs{\ln{B}}< 1$ indicate an \emph{inconclusive} evidence whereas values of ${\ln{B}}$ above 1, 2.5 and 5 indicate a \emph{weak}, \emph{moderate} and \emph{strong} evidence in favor of the \lcdm model, respectively.
%This is a conservative version of the so-called Jeffreys' Scale~\cite{Jeffreys1961}, as suggested by ref.~\cite{Trotta2008}.
%Note that $\ln{B} < -1$ means support in favor of the 2CC model.

% Priors for the LCDM model
%Finally, regarding the choice of the priors for the parameters of the \lcdm model, we followed the same methodology applied to obtain the priors for the 2CC parameters, i.e., using the results presented by ref.~\cite{CCCS} but now for the $\kd = \kg = k$ case.
%Therefore, we assumed the Gaussian priors $\OmegaMo = 0.277 \pm 0.017$ and $\Omegako = 0.0 \pm 0.006$ for the matter and curvature density parameters, respectively.
%The priors on $H_0$ and on the JLA nuisance parameters were kept the same as those shown in table~\ref{tab:priors}.

%Do for same data as used in previous 2-curvature model (Beethoven doing now).

%Discuss new results relative to 2-curvature model results.

\begin{table}[!t]
    \centering
    \caption{\label{tab:fixed_params}Fixed parameters for all the analyses (all values are from Planck's 2015 results \cite{planck2015}, except for $\Neff$).}
    \begin{tabular}{ll}
        \toprule
        Parameter  & Value \\
        \midrule
        $\OmegaBo$ & $0.041$ \\
        $\OmegaRo$ & $9.226 \times 10^{-5}$ \\
        $\zd$      & $1059.620$ \\
        $\Neff$    & $3.046$ \\
        \bottomrule
    \end{tabular}
\end{table}

\begin{table}[!t]
    \centering
    \caption{\label{tab:priors_to_H_analysis}Priors on the parameters of the 2CC-Ext and \lcdm models for the $H(z)$ analysis.}
    \begin{tabular}{ll}
        \toprule
        Parameter   & Prior \\
        \midrule
        $H_0$       & $\N{73.240}{3.028}$\footnotemark[1] \\
        $\OmegaMo$  & $\U{0}{1}$\footnotemark[2] \\
        $\OmegaKdo$ & $\U{-1}{1}$ \\
        $\OmegaKgo$ & $\U{-1}{1}$ \\
        $\alpha$    & $\U{0}{5}$ \\
        $\OmegaKo$  & $\U{-5}{5}$ \\
        \bottomrule
    \end{tabular}
\end{table}
\footnotetext[1]{$\N{\mu}{\sigma^2}$ is the Gaussian prior with mean $\mu$ and variance $\sigma^2$.}
\footnotetext[2]{$\U{a}{b}$ is the uniform prior.}

\begin{table}[!h]
    \centering
    \caption{\label{tab:priors_from_H_analysis}Priors on the parameters of the 2CC-Ext and \lcdm models for the JLA, BAO and JLA + BAO analyses. They are the results of the $H(z)$ analysis.}
    \begin{tabular}{lll}
        \toprule
        Parameter   & 2CC-Ext             & \lcdm \\
        \midrule
        $H_0$       & $\N{72.336}{1.527}$ & $\N{72.678}{1.639}$ \\
        $\OmegaMo$  & $\N{0.507}{0.238}$  & $\N{0.447}{0.151}$ \\
        $\OmegaKdo$ & $\N{-0.441}{0.362}$ & \dots \\
        $\OmegaKgo$ & $\N{0.023}{0.579}$  & \dots \\
        $\alpha$    & $\N{0.959}{0.159}$  & \dots \\
        $\OmegaKo$  & \dots               & $\N{-0.448}{0.360}$ \\
        \bottomrule
    \end{tabular}
\end{table}

\subsection{Data}

The data sets that we used are
\begin{itemize}
    \setlength{\itemsep}{-2.5pt}
    \item $H(z)$: Compilation of values of $H(z)$ provided in Ref.~\cite{Wang2017} using the differential age method.
    \item JLA: The SNe Ia compilation from Ref.~\cite{Betoule2014}.
    \item BAO: The $\theta_\mathrm{BAO}(z)$ data from Refs.~\cite{Carvalho2016, Alcaniz2017}.
\end{itemize}

In Table~\ref{tab:fixed_params}, the fixed cosmological parameters that we use for all the analyses are given. The parameters are from Planck's 2015 results, except for $N_\mathrm{eff}$. Table \ref{tab:priors_to_H_analysis} contains the Priors on the parameters of the 2CC-Ext and \lcdm models for the $H(z)$ analysis. Similarly, Table \ref{tab:priors_from_H_analysis} contains the Priors on the parameters of the 2CC-Ext and \lcdm models for the JLA, BAO and JLA + BAO analyses. These are derived results from the $H(z)$ analysis.

\section{Results} \label{results}

\begin{table}[!h]
    \centering
    \caption{\label{tab:results_parameterestimation}Results for the JLA, BAO and combined JLA + BAO analyses using the priors from Table~\ref{tab:priors_from_H_analysis}. The numbers read ``mean $\pm$ one standard deviation''.}
    \begin{tabular}{l d{2.11} d{2.11}}
        \toprule
        Parameter   & \multicolumn{1}{c}{2CC-Ext} & \multicolumn{1}{c}{\lcdm} \\
        \midrule

        \multicolumn{3}{c}{\rule{0pt}{1.5em}JLA\rule[-1em]{0pt}{0.5em}} \\
        $H_0$       & 72.310 \pm 1.386            & 72.675 \pm 1.479 \\
        $\OmegaMo$  & 0.562 \pm 0.159             & 0.331 \pm 0.074 \\
        $\OmegaKdo$ & -0.280 \pm 0.242            & \multicolumn{1}{c}{$\dots$} \\
        $\OmegaKgo$ & 0.300 \pm 0.384             & \multicolumn{1}{c}{$\dots$} \\
        $\alpha$    & 0.895 \pm 0.120             & \multicolumn{1}{c}{$\dots$} \\
        $\OmegaKo$  & \multicolumn{1}{c}{$\dots$} & -0.071 \pm 0.174 \\

        \multicolumn{3}{c}{\rule{0pt}{1.5em}BAO\rule[-1em]{0pt}{0.5em}} \\
        $H_0$       & 72.301 \pm 1.483            & 72.699 \pm 1.631 \\
        $\OmegaMo$  & 0.616 \pm 0.164             & 0.263 \pm 0.047 \\
        $\OmegaKdo$ & -0.314 \pm 0.298            & \multicolumn{1}{c}{$\dots$} \\
        $\OmegaKgo$ & -0.068 \pm 0.534            & \multicolumn{1}{c}{$\dots$} \\
        $\alpha$    & 0.985 \pm 0.006             & \multicolumn{1}{c}{$\dots$} \\
        $\OmegaKo$  & \multicolumn{1}{c}{$\dots$} & 0.192 \pm 0.207 \\

        \multicolumn{3}{c}{\rule{0pt}{1.5em}JLA + BAO\rule[-1em]{0pt}{0.5em}} \\
        $H_0$       & 72.335 \pm 1.396            & 72.674 \pm 1.517 \\
        $\OmegaMo$  & 0.549 \pm 0.144             & 0.249 \pm 0.018 \\
        $\OmegaKdo$ & -0.332 \pm 0.235            & \multicolumn{1}{c}{$\dots$} \\
        $\OmegaKgo$ & 0.446 \pm 0.354             & \multicolumn{1}{c}{$\dots$} \\
        $\alpha$    & 0.985 \pm 0.006             & \multicolumn{1}{c}{$\dots$} \\
        $\OmegaKo$  & \multicolumn{1}{c}{$\dots$} & 0.132 \pm 0.061 \\

        \bottomrule
    \end{tabular}
\end{table}

\begin{table}[!t]
    \centering
    \caption{\label{tab:results_modelselection}Bayesian evidence and Bayes factors for the 2CC-Ext model related to the \lcdm model. The last column shows the interpretation of the 2CC-Ext model's evidence compared to the evidence of the \lcdm model, following the Jeffreys' scale.}
    \begin{tabular}{l d{4.10} d{3.10} c}
        \toprule
        Model & \multicolumn{1}{c}{$\ln{\mathcal{E}}$} & \multicolumn{1}{c}{$\ln{B}$} & Evidence in favor of 2CC-Ext \\
        \midrule
        \multicolumn{4}{c}{\rule{0pt}{1.5em}JLA\rule[-1em]{0pt}{0.5em}} \\
        2CC-Ext & -354.981 \pm 0.389 & 1.062 \pm 0.390  & Inconclusive --- Weak (favored) \\
        \lcdm   & -356.043 \pm 0.024 & 0                & \dots \\
        \multicolumn{4}{c}{\rule{0pt}{1.5em}BAO\rule[-1em]{0pt}{0.5em}} \\
        \lcdm   & -9.478 \pm 0.025   & 0                & \dots \\
        2CC-Ext & -10.258 \pm 0.037  & -0.780 \pm 0.045 & Inconclusive \\
        \multicolumn{4}{c}{\rule{0pt}{1.5em}JLA + BAO\rule[-1em]{0pt}{0.5em}} \\
        \lcdm   & -362.885 \pm 0.073 & 0                & \dots \\
        2CC-Ext & -365.132 \pm 0.023 & -2.247 \pm 0.077 & Weak (disfavored) \\
        \bottomrule
    \end{tabular}
\end{table}

The main quantitative results of our Bayesian analysis can be found in Tables \ref{tab:results_parameterestimation} and~\ref{tab:results_modelselection}. In Table \ref{tab:results_parameterestimation}, we obtain the estimated parameters for the 2CC-Ext model, using JLA+BAO data. Here we note that the energy density of matter in such a model would be much larger than the standard model. The model also prefers a universe with non-zero spatial curvature. In addition, the dynamical $k_d$ and geometric curvature $k_g$ both differ. Also, differing values of the geometric and dynamical Hubble rates are also preferred, with the parameter $\alpha$ deviating slightly from one. This could suggest that such a model might resolve the $H_0$ tension. However, to be certain we would also need to include CMB data.

For model selection, Table \ref{tab:results_modelselection} tells us that The JLA supernovae data and the BAO data cannot independently distinguish between the 2CC-Ext model and the \lcdm model. However, when we combine both the JLA and BAO data we find $\ln{B} = -2.247 \pm 0.077$, which indicates that the evidence weakly favours the standard \lcdm model over the 2CC-Ext model. This is in stark contrast to the results of \cite{SantosColey}, which performed a similar analysis for the two spatial curvature model, single Hubble rate model i.e., ``the 2CC model''. In \cite{SantosColey}, the 2CC model was moderately favoured over the standard model. There could be two reasons for this discrepancy. Firstly, the additional parameter $\alpha$ in the 2CC-Ext for two Hubble scales has weakened the constraints on the model, as compared to the two curvature, single Hubble rate 2CC model. Secondly, we used updated priors from updated data for the 2CC-Ext model analysis, which could also have affected the final results.

%From these results, we observe that the current SNe Ia data alone cannot distinguish between the 2CC and standard models, with $\ln{B} = 0.884 \pm 3.865$, which varies from moderate evidence against to moderate evidence in favor of the \lcdm scenario.
%On the other hand, the BAO and the SNe Ia + BAO data are much more effective to this end, as can be seen in the second and third sub-tables of table~\ref{tab:results_modelselection}.
%For the joint analysis of SNe Ia and BAO measurements, we find $\ln{B} = -6.549 \pm 5.264$, which indicates that the evidence of the standard \lcdm model varies from strongly to moderately disfavored with respect to the 2CC cosmology.
%A graphical representation of the ranges of all Bayes factors is displayed in the figure~\ref{fig:logBFs}.
%It is worth mentioning that we also explored the dependence of the results on a different prior on $H_0$~\cite{planck2015} and we find that the results of table~\ref{tab:results_modelselection} remain unchanged.

%\subsection{Discussion}

%Do for new data [the data Beethoven sent to us before]. Discuss and compare results.
%[new subsection or put in final Discussion].

%\section{Discussion}

\section{Conclusions}

We do not necessarily present these results as evidence for or against the simple phenomenological model discussed here, but rather more as motivation for studying more physical models that take into consideration the possible effects of backreaction which can affect cosmological observations at the level of a few percent. In the era of precision cosmology percentage level effects should be considered significant and we might need more sophisticated models beyond the standard model to understand them.
%{\bf{MORE}}

In this paper, we review the $H_0$ problem in cosmology. We also review the role of spatial curvature in precision cosmology and how it might might be tied in with the $H_0$ problem and our understanding beyond the standard model. In addition, we present certain physically motivated models that fit under an extended phenomenological template that can be compared to the standard \lcdm model. Finally, we perform a Bayesian analysis to estimate parameters in this extended phenomenological model, and also compute the Bayesian Evidence to do a model comparison between the 2CC-Ext model and standard model. The main results from late time JLA+BAO data tell us that the standard model is weakly preferred, suggesting that studying models that fit within the 2CC-Ext umbrella might be futile. However, this can be confirmed with greater certainty, once we can include the updated Planck CMB data \cite{planck2018}. If there is any indication from CMB data that these models might be interesting to study, these 2CC-Ext models might also be able to resolve the $H_0$ tension.

\section*{Acknowledgements} AC acknowledges the support of the NSERC and VAAS acknowledges the support of AARMS. BS is supported by the National Observatory DTI-PCI program of the Brazilian Ministry of Science, Technology, and Innovation (MCTI). We would also like to thank David Wiltshire, Thomas Buchert and Jailson Alcaniz for helpful comments.

%%%%%%%%%%%%%%%%%%%%%%%%%%%%%%%%%%%%%%%%%%%%%%%%%%%%%%%%%%%%%%%%%%%%%%%%%%%%%%%
\bibliographystyle{JHEP}

\providecommand{\href}[2]{#2}
\begingroup
\raggedright

\endgroup
%%%%%%%%%%%%%%%%%%%%%%%%%%%%%%%%%%%%%%%%%%%%%%%%%%%%%%%%%%%%%%%%%%%%%%%%%%%%%%%

\end{document}